\documentclass[amssymb,prb,aps,twocolumn,amsmath,showpacs]{revtex4}

\usepackage[dvips]{graphicx}
\usepackage{amstext}
\usepackage{textcomp}

\begin{document}

\title{Spin-valve Josephson junctions for cryogenic memory}
\author{Bethany M. Niedzielski}\altaffiliation[Present address: ]{MIT Lincoln Labs, 244 Wood St, Lexington, MA 02421}
\author{T.J. Bertus}
\author{Joseph A. Glick}
\author{R. Loloee}
\author{W. P. Pratt, Jr.}
\author{Norman O. Birge}
\email{birge@pa.msu.edu} \affiliation{Department of Physics and
Astronomy, Michigan State University, East Lansing, Michigan
48824-2320, USA}

\date{\today}

\begin{abstract}

Josephson junctions containing two ferromagnetic layers are being considered for use in cryogenic memory.  Our group recently demonstrated that the ground-state phase difference across such a junction with carefully chosen layer thicknesses could be controllably toggled between zero and $\pi$ by switching the relative magnetization directions of the two layers between the antiparallel and parallel configurations.  However, several technological issues must be addressed before those junctions can be used in a large-scale memory.  Many of these issues can be more easily studied in single junctions, rather than in the Superconducting QUantum Interference Device (SQUID) used for the phase-sensitive measurements.  In this work, we report a comprehensive study of spin-valve junctions containing a Ni layer with a fixed thickness of 2.0 nm, and a NiFe layer of thickness varying between 1.1 and 1.8 nm in steps of 0.1 nm.  We extract the field shift of the Fraunhofer patterns and the critical currents of the junctions in the parallel and antiparallel magnetic states, as well as the switching fields of both magnetic layers.  We also report a partial study of similar junctions containing a slightly thinner Ni layer of 1.6 nm and the same range of NiFe thicknesses.  These results represent the first step toward mapping out a ``phase diagram" for phase-controllable spin-valve Josephson junctions as a function of the two magnetic layer thicknesses.

\end{abstract}

 \maketitle

\section{Introduction}

Experimental studies of Josephson junctions containing ferromagnetic layers blossomed after the seminal work of the Ryazanov and Aprili groups in 2001 and 2002 \cite{Ryazanov:01,Kontos:02}.  Those workers were the first to confirm the theoretical prediction \cite{Buzdin:82, BuzdinReview} that the ground-state phase difference across an S/F/S Josephson junction could be either zero or $\pi$, depending on the thickness of the ferromagnetic F-layer inside the junction.  Oscillations in the magnitude of the critical current as a function of F-layer thickness, signifying oscillations between 0 and $\pi$ junctions, have now been demonstrated with a wide variety of ferromagnetic materials, including CuNi alloy,\cite{Ryazanov:01,Sellier:03,Oboznov:06,Weides:06} PdNi, \cite{Kontos:02, Khaire:09} Ni,\cite{Blum:02,Shelukhin:06,Robinson:06, Bannykh:09, Baek:14, Baek:17} Ni$_3$Al,\cite{Born:06} Co,\cite{Robinson:06} Fe,\cite{Robinson:06} NiFe,\cite{Robinson:06,Qadar:14,Glick:17} NiFeMo,\cite{Niedzielski:15} PdFe,\cite{GlickPdFe:17} and NiFeCo.\cite{Glick:17}

More interesting still is the possibility of controlling either the amplitude of the critical current or the ground-state phase difference across the junction by inserting two different ferromagnetic layers and controlling the relative orientation of their magnetizations.\cite{Krivoruchko:01, Golubov:02,Bell:04,Pajovic:06} This could be accomplished by using a ``pseudo spin valve" consisting of a magnetically ``hard" material for the fixed layer and a magnetically ``soft" material for the free layer, so that the magnetization of the free layer can be reversed by a small magnetic field without disturbing the magnetization of the fixed layer.  This creates two magnetic states when the layer magnetizations are parallel or antiparallel which can have different critical current or phase shift values.  As early as 2004, Bell \textit{et al.}\cite{Bell:04} suggested that such a spin-valve junction could be used as a cryogenic memory element, and that idea is now being actively pursued by several groups.\cite{Baek:14,Qadar:14,Niedzielski:15,Gingrich:16,Dayton:17}  The memory design that our group is working towards was proposed by workers at Northrop Grumman Corporation several years ago.\cite{HerrPatent1}  It was soon realized, however, that basing a memory cell on the critical current of the spin-valve junction would limit the read speed of the memory, due to the rather low values of $I_cR_N$ typical of such junctions,\cite{Larkin:12} where $I_c$ is the critical current and $R_N$ is the resistance in the voltage state.  The Northrop Grumman team responded to that challenge by suggesting to use the phase state of the controllable junction as the memory storage element.\cite{HerrPatent2}  If the controllable spin-valve junction is inserted into a SQUID loop containing two conventional S/I/S junctions with $I_c$'s that are smaller than that of the controllable junction, then only the S/I/S junctions will switch into the voltage state during a read operation and determine the read speed, while the controllable junction always stays in the supercurrent state \cite{Dayton:17}.  That memory design motivated our recent experimental demonstration of controllable switching of a spin-valve junction between the 0 and $\pi$ states.\cite{Gingrich:16}

 Demonstration of controllable 0 - $\pi$ switching in a single device is a first step toward fabricating the Northrop Grumman memory, but there is a long way to go to implement the technology on a large scale.  The spin-valve junctions in our 0 -- $\pi$ demonstration suffered from several drawbacks: 1) The very thin Ni fixed layers required rather large external fields -- of order 220 mT -- to saturate the magnetization and initialize the memory bit; 2) Device behavior varied somewhat on different cooldowns, suggesting that the Ni layers had inhomogeneous magnetization -- likely a multi-domain state -- even after initialization in a large field; 3) The switching fields of the NiFe free layers varied somewhat from device to device, probably due to disorder and roughness in the surrounding layers; and 4) The critical current when the magnetizations of the two layers were antiparallel to one another (AP state) was substantially larger than when the magnetizations of the two layer were parallel (P state) for all samples studied.  Since that work was completed, we have improved the smoothness of both our Nb base electrode and the Cu interlayers surrounding the ferromagnetic layers.  We have also experimented with slightly thicker Ni layers than in our previous work, since the magnetic behavior of thin magnetic films tends to degrade as their thickness is reduced.  Before carrying out additional phase-sensitive measurements, it is advantageous to first measure the properties of single Josephson junctions to avoid the complications of the phase-sensitive SQUID measurements.  In this paper we report the complete characterization of spin-valve junctions containing a Ni layer with fixed thickness of 2.0 nm as the magnetic hard layer, and a NiFe layer with thickness varying between 1.1 and 1.8 nm as the magnetic soft layer.  From these measurements, we obtain the critical currents in both the P and AP states as a function of NiFe thickness, and the switching field ranges for both magnetic layers.  In addition, we discuss the field shifts of the ``Fraunhofer" patterns due to the intrinsic magnetization of both layers.  We also report critical currents of a second set of samples containing a Ni layer with fixed thickness of 1.6 nm and a NiFe layer with the same thickness range as in the first set. Together, these two data sets provide the first step toward mapping out the phase diagram of phase-controllable spin-valve Josephson junctions containing Ni and NiFe ferromagnetic layers, as a function of the layer thicknesses.

 A similar study on Ni/NiFeNb spin-valve junctions was carried out by Baek \textit{et al.} a few years ago.\cite{Baek:14}  The work reported here uses NiFe rather than NiFeNb as the soft magnetic layer, and we include additional kinds of measurements and data analysis not presented in that earlier study.

\section{Sample Fabrication}

A schematic cross-section of the Josephson junction samples is shown in Fig. 1.  For the bottom superconducting electrode we use a [Nb/Al] multilayer rather than pure Nb, since the former has considerably less roughness \cite{Kohlstedt:96,Thomas:98,Wang:12}.  The samples were fabricated in three main steps: i) sputtering and large-scale patterning of a multilayer stack of the form: [Nb(25)/Al(2.4)]$_3$/Nb(20)/Cu(2)/NiFe($d_\mathrm{NiFe}$)/Cu(4)/
Ni(2)/Cu(2)/Nb(5)/Au(15), with all thicknesses in nm and the subscript giving the number of repeats; ii) patterning of elliptical Josephson junctions with dimensions 1.25$\mu$m $\times$ 0.5$\mu$m by electron-beam lithography and Ar ion milling, followed by deposition of a 50-nm SiO$_x$ insulating layer, then lift off of the e-beam resist; and iii) sputtering of the thick top Nb electrode, Nb(150)/Au(10), through a photolithography stencil after a brief ion mill to clean any resist residue from the top Au layer of the previous step.  Sputtering was performed with the samples cooled to between -30\textdegree ~C and -15\textdegree ~C in an Ar pressure of about 2 mTorr.  The base pressure of the sputtering system was less than 2 $\times$ $10^{-8}$ Torr.  The e-beam lithography was performed using the negative e-beam resist, ma-N2401. The large-scale patterning in the first and last fabrication steps were performed using photolithography with S1813 resist treated with chlorobenzene to form a partial undercut and ease the lift-off process.

\begin{figure}[tbh!]
\begin{center}
\includegraphics[width=2.5in]{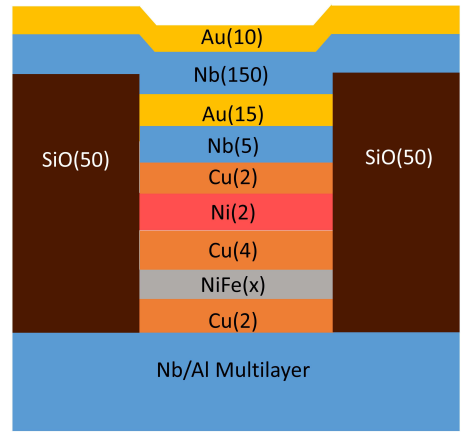}
\end{center}
\caption{Schematic cross-section of the Josephson junction samples, with all thicknesses in nm.  The diagram is not to scale.  In the second set of junctions discussed later, the Ni thickness is 1.6 nm rather than 2.0 nm.}
\label{SampleStructure}
\end{figure}

\section{Experimental Results}

Two Josephson junctions from the same chip were measured for each NiFe thickness in the series.  All measurements in this study were performed at 4.2 K, with the sample immersed in liquid helium.  Measurements of current-voltage characteristics (I-V curves) were performed using a battery-operated current source and a SQUID-based self-balancing potentiometer with a voltage noise of a few pV/$\sqrt{Hz}$.\cite{Edmunds:80, Glick:17}  As shown in the inset to Fig. 2(a), the SQUID-based potentiometer has a voltage limit of less than 1$\mu V$; hence the full shapes of the I-V curves for $I > I_c$ are not visible for junctions with large values of critical current ($I_c$) and normal-state resistance ($R_N$), such as those in this study.  Fortunately, we know from measurements on similar junctions using conventional electronics\cite{Glick:17} that the I-V curves follow the standard square-root form for overdamped Josephson junctions.\cite{Barone:82,IVcurvenote}  To obtain a reliable value of $R_N$ using only the SQUID-based potentiometer, we measure the junctions at high magnetic field where $I_c$ = 0 and the I-V curve is Ohmic, as shown in the inset of Fig. 2(b).  (Note the very different current scales in the two insets.) The value of $R_N$ obtained from the slope, 22 $m\Omega$ for this junction, is used in the square-root fits to the I-V curves shown in the Fig. 2(a) inset, but the slopes are so high for $I > I_c$ that the value of $R_N$ used has negligible influence on the value of $I_c$ obtained from the fits.

Before any measurements are made, the sample is subjected to a large magnetic field -- typically 150 mT -- along the junction's long axis.  That initialization field is large enough to align the magnetizations of the both the Ni and NiFe layers in the junction.  Characterization of the samples involved several different types of measurements.  The conventional ``Fraunhofer pattern" of the junctions is obtained by sweeping the magnetic field slowly between -60 and +60 mT and then back again.  (The sample is initialized with a field of -150 mT before the upsweep, and +150 mT before the downsweep.)  Data for the sample with $d_\mathrm{NiFe}$ = 1.7 nm are shown in Fig. 2(a) and 2(b), for the upsweep and downsweep, respectively.  In both data sets, the data exhibit a large jump in the vicinity of zero field.  This is a signal that the direction of the NiFe magnetization is changing, and that the junction is thereby transitioning from the parallel (P) magnetization state to the antiparallel (AP) state.  There are two consequences of this transition.  First, the overall magnitude of the critical current changes; we will discuss that issue later.  Second, the horizontal shift of the pattern changes due to the change in the net magnetization of the junction from the Ni and NiFe layers.  The latter issue is well understood, and has been emphasized already in several previous works.\cite{Ryazanov:99, Khaire:09, Vernik:13, Baek:14, Niedzielski:15, Glick:17}  We include analysis of this issue here as a way of checking the consistency of our Josephson junction data.

For elliptical junctions, the form of the Fraunhofer pattern is:
\begin{equation}
	\label{eqn:FraunhoferAiryFit}
	I_{c}=I_{c0} \left| 2 J_{1} \left( \pi \Phi / \Phi_{0} \right) / \left( \pi \Phi / \Phi_0 \right) \right|,
	\end{equation}
where $I_{c0}$ is the maximum critical current, $J_1$ is a Bessel function of the first kind, $\Phi_0 = h/2e$ is the flux quantum, and $\Phi$ is the flux through the junction. If the external field $H$ is applied along the major axis of the ellipse and the magnetizations of the F layers are uniform and collinear with $H$, then the flux through the junction can be written as:
	\begin{equation}
	\label{eqn:magneticflux}
	\Phi=\mu_0 H w (2 \lambda_L+d_N+d_{F1} + d_{F2}) + \mu_0 w (M_1 d_{F1} + M_2 d_{F2}),
	\end{equation}
where $w$, $\lambda_L$, $d_N$, $d_{F1}$ and $d_{F2}$ are the width of the junction (minor axis), the London penetration depth of the Nb electrodes, the total thickness of all the normal metal layers, and the thicknesses of the two individual F layers, respectively.  (Eqn.~(\ref{eqn:magneticflux}) neglects the small demagnetizing field and any magnetic flux from the F layers that returns inside the junction.) From Eqn.~(\ref{eqn:magneticflux}) it is clear that the Fraunhofer pattern will be shifted along the field axis by an amount $H_{\mathrm{shift}}= -(M_1 d_{F1} + M_2 d_{F2})/(2 \lambda_L+d_N+d_{F1} + d_{F2})$ in the opposite direction of the junction's total magnetization.  This shift is expected to be largest in the P state when the magnetizations $M_1$ and $M_2$ have the same sign, and smaller in the AP state when they have opposite signs.

To obtain the values of $I_{c0}$ and $H_{\mathrm{shift}}$ in the P and AP states, we fit Eqn.~(\ref{eqn:FraunhoferAiryFit}) to the separate data for the P and AP states.  In the fitting procedure, the junction width $w$ is kept fixed at its nominal value of 0.5$\mu$m, and $\lambda_L$ is fixed at the value 85 nm determined from measurements on similar junctions over many years.\cite{Khaire:09} Figures 2(a) and 2(b) show fits to the P and AP states as solid black and red lines, respectively.  The points used in the fits are represented by solid symbols, while the points in the transition region are indicated by hollow symbols.  Unfortunately, the field range over which we have data for the AP state is rather limited.  While the NiFe magnetization switches direction abruptly at low field, ~ 2 - 4 mT, the Ni magnetization switches direction gradually over a range typically 25 - 100 mT.  Hence we limited the field range for the fits to the AP state data to a maximum value in the vicinity of 25 mT for all samples.  (It may be possible in some cases to fit Fraunhofer patterns even when the magnetization is changing with field \cite{Golovchanskiy:16}, but we prefer to avoid the complications involved in attempting such fits.)  For the P state, we fit the data over a broad field range starting from the beginning of the sweep and ending just shy of zero field.  Because of the large shifts, however, the robustness of the fits in the P state suffer from the fact that the central peaks of the Fraunhofer patterns are usually far outside this range.  In Figures 2(a) and 2(b) we extend the solid lines beyond the range of the data used in the fits, so the reader can see the positions of the central peaks identified by the fits.  The values of $H_{shift}$ are labeled with arrows for both data sets in both figures.

\begin{figure}[tbh!]
\begin{center}
\includegraphics[width=3.2in]{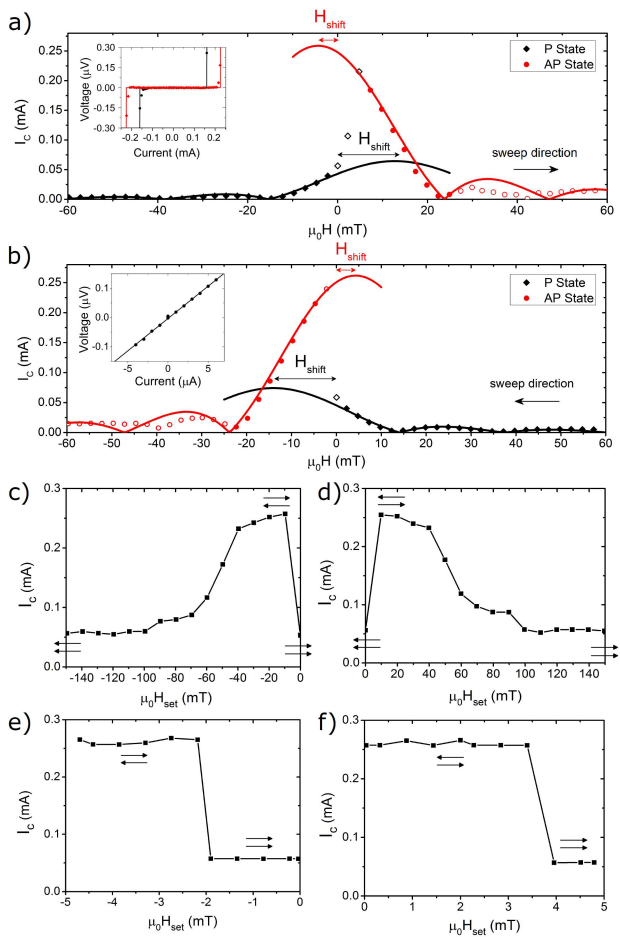}
\end{center}
\caption{Characterization of a Josephson junction containing a 2.0-nm Ni fixed layer and a 1.7-nm NiFe free layer. a) Critical current vs field (the Fraunhofer pattern) obtained with the field sweeping from left to right after application of an initialization field of -150 mT.  Solid black and red lines show fits to the data in the P and AP states, respectively.  Only the data points represented by solid symbols are used in the fits.  Inset: I-V curves measured at fields of +5 and +10 mT during the upsweep. Solid lines are fits discussed in the text.  b) Similar data and fits for the field sweep from right to left after application of an initialization field of +150 mT.  The field shifts of the Fraunhofer patterns are labeled as $H_{shift}$.  Inset: I-V curve measured at a field of -120 mT, where $I_c$=0. (Note the very different current scales in the two insets.) The slope of the fit line provides the normal-state resistance, $R_N$ = 22 m$\Omega$ for this junction.  c)-f) Critical current measured at zero field vs set field.  Panels c) and d) show ``major loop" data, while panels e) and f) show ``minor loop" data. c) After initialization with a positive field, the set field is sequentially increased in the negative direction, showing switching first of the NiFe at very low field followed by the Ni at larger field.  d) Data taken after those in c), with the set field now increasing in the positive direction to return the sample to the initial state.  e) Similar to c), but with the field sweep stopping before the Ni fixed layer switches. f) Following e), the sample is returned to its initial state.  }
\label{Fraunhofer}
\end{figure}

In the Fraunhofer data, the effect of the magnetization switching is intertwined with the Fraunhofer physics.  While it is fairly easy to point out the switching of the NiFe magnetization, it is not at all clear where the Ni magnetization is switching.  Hence we carried out another type of measurement, performed at zero field, shown in Figures 2(c) and 2(d).  These measurements involve stepping a ``set field", $H_{set}$, through a sequence of values, and measuring the I-V curve after each step.  This type of measurement is directly relevant to real devices used in memory cells, which are likely to be ``read" in zero field. For these data, the sample was first initialized in a large field of 150 mT in the positive direction. The set field was then stepped in the negative direction with a step size of 10 mT, and an I-V curve was measured at zero field after each step.   Fig. 2(c) shows a sudden jump up in $I_c$ at the first step, indicating switching of the NiFe free layer magnetization, and then a gradual return of the original value of $I_c$ as the set field stepped from -40 to -100 mT, indicating reversal of the Ni fixed layer magnetization.  This allows us to determine the saturation field for the Ni layer.  Fig. 2(d) shows the reverse process when the field is stepped in the positive direction.  Sets of double arrows on both plots indicate the magnetization directions of the Ni layer on top and the NiFe layer on the bottom.

In a practical application, the magnetic field used to ``write" the device will always be kept low enough so that only the free layer magnetization switches, while the fixed layer stays fixed.  Fig. 2(e) and 2(f) show high-resolution switching data for the free layer only -- so-called ``minor loops" in contrast to the ``major loops" shown in Fig. 2(c) and 2(d).  The field step size here was about 0.5 mT, although the points do not lie exactly on multiples of 0.5 mT due to the limited resolution of the magnet power supply.  The plot in Fig. 2(e) shows the NiFe magnetization switching close to -2 mT, while Fig. 2(f) shows it switching back to the initialized direction at +4 mT.  We expect the magnitude of the switching field from the AP to P state to be somewhat larger than that from the P to AP state, due to magnetostatic dipolar coupling between the Ni and NiFe layers.  This sample follows that expectation, but not all samples do.  Departures from the expected behavior are probably due to disorder and roughness in the NiFe layer, as well as a multi-domain state in the Ni layer with corresponding non-uniform magnetization.

\section{Data Analysis and Discussion}

The primary motivation for carrying out this study was to compare the critical currents in the P and AP states, with the goal of determining where these samples lie on a phase diagram of 0 and $\pi$ junctions vs Ni and NiFe thicknesses.  Creating a definitive phase diagram requires performing phase-sensitive measurements; nevertheless, the presence of 0 - $\pi$ transitions in the diagram is signalled in single-junction measurements by zeroes (or deep minima) in the magnitude of $I_c$.  Fig. 3 shows two plots of $I_cR_N$ vs NiFe layer thickness, $d_\mathrm{NiFe}$, for both the P and AP states.  Fig. 3(a) uses the values of $I_c$ obtained from the raw data at zero field, such as those shown in Figs. 2(c-f).  Fig. 3(b) uses the values of $I_c$ obtained from the fits to the Fraunhofer patterns, such as those shown in Figs. 2(a) and 2(b).  The scatter in the values of $I_cR_N$ for two junctions on the same chip is typical for junctions containing very thin layers of strong ferromagnetic materials such as NiFe \cite{Glick:17}.  The data in Fig. 3 strongly suggest that there is a zero of $I_cR_N$ in the P state -- hence a 0 - $\pi$ transition -- when $d_\mathrm{NiFe} \approx 1.5$nm, while the AP state exhibits a zero and a corresponding 0 - $\pi$ transition when $d_\mathrm{NiFe} \approx 1.0$nm.  Those observations are the primary result of this work.

\begin{figure}[tbh!]
\begin{center}
\includegraphics[width=3.0in]{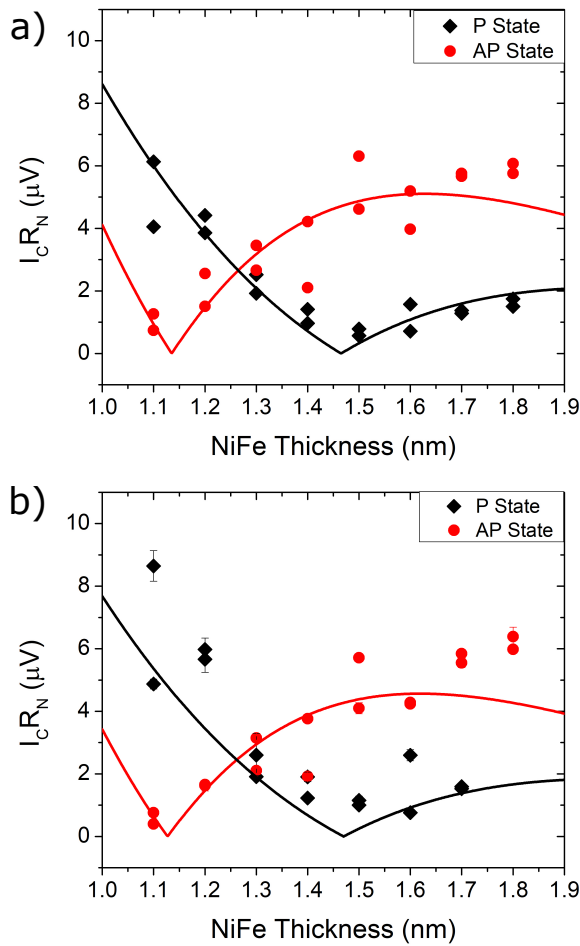}
\end{center}
\caption{(color online) Critical current times normal state resistance for all Josephson junctions with $d_\mathrm{Ni}$=2.0 nm, vs NiFe layer thickness.  Red circles represent the AP state while black diamonds represent the P state.  a) Critical current obtained from zero-field data, as shown in Figs. 2c) - 2f). b) Critical current obtained from fits to Fraunhofer patterns such as those shown in Fig. 2a) and 2b).   The solid lines in a) and b) are fits to Eqns. (17) and (20) in Ref. \onlinecite{Crouzy:07} as described in the text.  Error bars in b) come from the Fraunhofer fitting procedure, and reflect the relative quality of those fits.  Those error bars are used to weight the points in the fitting procedure, but they are much smaller than the sample-to-sample variations seen in the data.  There are no error bars in a) and all data points are given equal weight in the fits.}\label{IcRn}
\end{figure}

Comparing the two data sets, one might expect Fig. 3(b) to represent the ``true" value of $I_cR_N$, since it displays the expected peak value in cases where the peak is outside the range of the data.  We have found, however, that Fraunhofer patterns of junctions containing only a single magnetic layer do not always follow the theoretical form exactly.\cite{Niedzielski:15, Glick:17}  We occasionally observe patterns where the heights of the side lobes are larger than expected relative to the height of the central peak.  In such cases, the fit of Eqn.~(\ref{eqn:FraunhoferAiryFit}) to the data may exaggerate the height of the central peak.  Junctions with  a larger shift in the Fraunhofer pattern are more likely to be affected by this since there are limited data in the central lobe for fitting.  Therefore, we maintain a certain level of healthy skepticism about the data presented in Fig. 3(b).  The data in Fig. 3(a), on the other hand, represent lower bounds to all values of $I_c$.  And, as noted earlier, zero-field measurements are more relevant to the read operation of a real memory device.  Fortunately, the differences between the two data sets are rather minor, except for when $d_\mathrm{NiFe}$ = 1.1 or 1.2 nm.

Before discussing the solid curves in Fig. 3, we complete the initial data analysis by discussing the field shifts of the Fraunhofer patterns, obtained from the fits shown in Figs. 2a) and 2b).  Fig. 4 shows the field shifts for all the samples measured, as a function of NiFe thickness.  The field shifts have opposite signs for the data in the two sweep directions, as is apparent in Figs 2a) and 2b).  To avoid clutter in Fig. 4, we take the average of the up- and down-sweep field shifts (while inverting the sign of the down-sweep shift), and plot the average with the sign appropriate for the up-sweep shift.  Since the sample was initialized with a large field in the negative direction before the up-sweep, both the Ni and NiFe layers in the junction have magnetizations pointing in the negative direction, so the field shift in the P state should be large and positive.  In addition, we expect the magnitude of the field shift to increase with NiFe thickness.  The black data points in Fig. 4 largely follow those expectations, with the exception of a few outlier points -- especially the data points for the sample with $d_\mathrm{NiFe}$ = 1.3 nm.  All junctions on that chip displayed anomalous field shifts, for which we do not know the origin.\cite{badmilling}  Fortunately, the critical currents of those samples are not anomalous, and follow the trends in Fig. 3 nicely.  When the sample switches to the AP state, the NiFe magnetization now points in the positive direction, so the magnetizations of the Ni and NiFe layers tend to cancel each other.  For the junctions with the thinnest NiFe, the Ni contribution to the shift is larger and the net shift is expected to be positive, while for the junctions with thickest NiFe, the NiFe contribution is larger and the net shift should be negative.  The red data points in Fig. 4 follow those expectations.

\begin{figure}[tbh!]
\begin{center}
\includegraphics[width=3.2in]{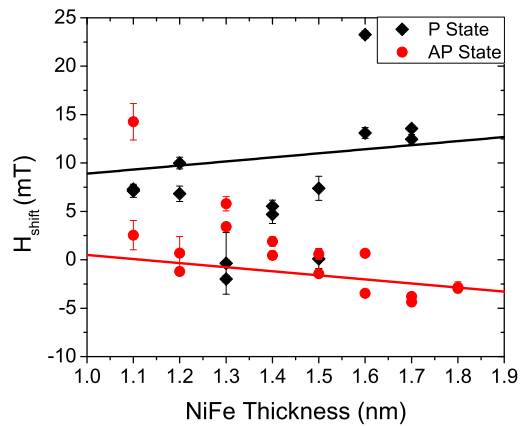}
\end{center}
\caption{(color online) Field shifts of Fraunhofer patterns for all Josephson junctions with $d_\mathrm{Ni}$=2.0 nm, vs NiFe layer thickness.  The data are obtained from the fits to the Fraunhofer patterns, as shown for one sample in Figs. 2a) and 2b).  The black and red lines are obtained from a simultaneous fit of Eqns. 3 and 4 to the two data sets.  As in Fig. 3, the error bars on the data points come from the individual fits to the Fraunhofer data, but they do not account for the larger sample-to-sample variations seen in the data.}\label{FieldShifts}
\end{figure}

Given the large scatter in the data shown in Fig. 4, performing unconstrained least-squares fits to each data set would likely return unphysical fit parameters.  Instead, we enforce the following contraint.  We recently reported a study of S/F/S Josephson junctions containing only a single NiFe layer \cite{Glick:17}; the Fraunhofer patterns of those junctions exhibit a field shift that increases linearly with NiFe thickness, with a slope of 5.0 mT/nm, and with a negligibly small intercept, indicating that any ``dead" magnetic layer at the Cu/NiFe interface is very small.  Similar measurements of S/F/S junctions containing a single Ni layer have been performed recently by Baek \textit{et al.}\cite{Baek:17}; unfortunately those authors only showed the field shift data for one sample.  If we neglect magnetic dead layers, then following Eqn.~(\ref{eqn:magneticflux}) we expect the field shifts in the P and AP states to follow:
    \begin{equation}
	\label{eqn:fieldshiftsP}
	H_{shift}^{P}=a d_\mathrm{NiFe} + b d_\mathrm{Ni}
    \end{equation}
    \begin{equation}
	\label{eqn:fieldshiftsAP}
	H_{shift}^{AP}=-a d_\mathrm{NiFe} + b d_\mathrm{Ni},
    \end{equation}
where \textit{a} and \textit{b} are constants approximately equal to, respectively, the magnetizations of NiFe and Ni divided by about 180 nm -- i.e. twice the London penetration depth of 85 nm plus the total thickness of the non-superconducting layers in the junctions.  Since the Ni thickness is fixed at 2.0 nm, the data for the two field shifts vs $d_\mathrm{NiFe}$ should lie on two straight lines with opposite slopes and a common y-intercept.  We carried out a simultaneous least-squares fit of Eqns.~(\ref{eqn:fieldshiftsP}) and (\ref{eqn:fieldshiftsAP}) to the P and AP state data, resulting in the black and red lines shown in Fig. 4. Those lines correspond to the values of the fit parameters: a = 4.25 mT/nm and b = 2.34 mT/nm.  The value of \textit{a} is only 15\% smaller than the value of 5.0 mT/nm obtained in Ref. \onlinecite{Glick:17}, while the latter is about 15\% smaller than we would estimate by using the Ni thin film magnetization data shown in Fig. 1(a) of Ref. \onlinecite{Baek:17}. There are a few outlier data points in Fig. 4, some of which we have already commented on.  In junctions where $I_cR_N$ is very small, it is difficult to obtain an accurate value of $H_{shift}$.  Examples include the very large value of $H_{shift}^{AP}$ for one sample with $d_\mathrm{NiFe}$ = 1.1 nm, the very small value of $H_{shift}^{P}$ in one sample with $d_\mathrm{NiFe}$ = 1.5 nm, and the very large value of $H_{shift}^{P}$ in one sample with $d_\mathrm{NiFe}$ = 1.6.  In all three of those cases, $I_cR_N$ is very small.  Those outliers do not affect the fit much because of the constraint that the two fit lines have the same intercept.  With the aforementioned caveats, we believe that the overall qualitative agreement between the field shift data and the two solid lines indicates that the Josephson junction samples are for the most part behaving magnetically as expected.

We now return to further discussion of Fig. 3.  Can we learn anything by fitting the data in Fig. 3 to a theoretical model?  In particular, will such a fit help us determine where these samples lie on a global phase diagram of 0 and $\pi$ states?  There are at least four theoretical papers that describe the critical current through spin-valve junctions of the form S/F/F/S or S/F/N/F/S.\cite{Blanter:04,Crouzy:07, Robinson:10, Melnikov:12}  (Several additional works address S/F/I/F/S junctions,\cite{Krivoruchko:01, Bergeret:01a,Barash:02,Golubov:02} but those calculations are not applicable to our samples.)  Before discussing which of the four relevant theories we should compare with our data, let us first discuss some of the general features of theories describing ballistic or diffusive transport.  An S/F/S Josephson junction with high-transparency interfaces and purely ballistic transport through F exhibits two pertinent features, due to the fact that electron trajectories at all angles propagate through F without scattering:\cite{Buzdin:82} i) the critical current decays only algebraically with increasing $d_F$ rather than exponentially; and ii) the first 0 -- $\pi$ transition occurs at a thickness $d_F/\xi_F \approx \pi/4$, with $\xi_F$ = $\hbar v_F/ 2 E_{ex}$ the coherence length of electron pairs in a ferromagnet (sometimes called the ``ferromagnetic coherence length" or just the ``magnetic length") in the ballistic limit.  (In the expression for $\xi_F$, $v_F$ and $E_{ex}$ are the Fermi velocity and exchange energy of the F material, respectively.) In a ballistic S$_1$/F$_1$/F$_2$/S$_2$ junction in the AP state, Blanter and Hekking\cite{Blanter:04} showed that, for every trajectory from S$_1$ to S$_2$, the center-of-mass phase accumulated by an electron pair traversing F$_1$ is partially canceled by the phase accumulated in F$_2$.  The result is that \textit{both} the amplitude of the critical current and the phase state of the junction depend only on the difference in thicknesses, $d_{F1} - d_{F2}$. (See Eqns. (9) and (10) in Ref. \onlinecite{Blanter:04} or Eqn. (6) in Ref. \onlinecite{Melnikov:12}.)  In the case of equal thicknesses, the junction behaves as though it contains no ferromagnetic material at all!  (This remarkable property disappears in the presence of disorder.\cite{Blanter:04})  In the P-state, not surprisingly, the junction behaves as though it contains a single ferromagnetic layer of thickness $d_{F1} + d_{F2}$.  Such a purely ballistic theory cannot fit the data in Fig. 3. As the NiFe thickness increases toward 1.8 nm, that theory would predict a very large magnitude of $I_cR_N$ in the AP state due to near cancellation of the pair phase accumulation in the Ni and NiFe layers.  In contrast, our samples show comparable magnitudes of $I_cR_N$ in the P and AP states as the NiFe thickness moves away from the locations of the $I_cR_N$ minima at $d_\mathrm{NiFe} \approx 1.5$nm in the P state and $d_\mathrm{NiFe} \approx 1.1$nm in the AP state.

A step away from a model of purely ballistic transport is a ``semi-ballistic" model for S/F/N/F/S junctions by Robinson \textit{et al.},\cite{Robinson:10} which incorporates both the mean free paths of the F and N materials and assumes that the Fermi surface of S is smaller than that of F so that the electron trajectories are directed mostly perpendicular to the interfaces.  Due to the latter assumption, the location of the first 0 -- $\pi$ transition in that model (in the P state) occurs at $d_F/\xi_F \approx \pi/2$, which is the result one finds in a purely one-dimensional model.  More importantly for us, $I_cR_N$ in both the P and AP states decays exponentially with the layer thicknesses, governed by the mean free paths in F and N.  Hence this model might, in principle, be applicable to our samples.  Unfortunately, those authors provide an explicit formula for $I_cR_N$ in the AP state that is valid only for equal thicknesses of the two F layers, so we cannot apply their model to our data.

A theory for diffusive S/F$_1$/N/F$_2$/S junctions in the P and AP states was provided by Crouzy, Tollis and Ivanov,\cite{Crouzy:07} and is valid for arbitrary values of the thicknesses $d_{F1}$ and $d_{F2}$.  The Crouzy theory has several limitations: it assumes rigid boundary conditions for the superconducting order parameter, and it does not take into account finite transparency of the interfaces, the finite mean free path of electrons, or spin-flip and spin-orbit scattering.  And like all calculations based on the Usadel equations, the theory cannot incorporate the complex band structure of strong ferromagnetic materials such as Ni and NiFe.  Nevertheless, we will show below that the Crouzy theory reproduces the most important qualitative features of our data.  Two salient features of the Crouzy theory are: i) the position of the first 0 -- $\pi$ transition in the P state occurs when $(d_{F1} + d_{F2})/\xi_F \approx 3\pi/4$; and ii) $I_cR_N$ decays exponentially with thickness in both the P and AP states according to exp$(-(d_{F1} + d_{F2})/\xi_F))$.  (Note that the definition of $\xi_F$ in the diffusive limit is $\xi_F$ = ($\hbar D/E_{ex})^{1/2}$ with $D$ the diffusion constant in F.)  The first of those results is consistent with previous theories of diffusive S/F/S junctions,\cite{BuzdinReview} although it has been noted by Faure \textit{et al}\cite{Faure:06} and by Heim \textit{et al}\cite{Heim:15} that the location of the first 0 -- $\pi$ transition can vary widely if there are tunnel barriers or extra non-magnetic layers in the junction.

We fit Eqns. (17) and (20) in Ref. \onlinecite{Crouzy:07} to our data for the P and AP states, respectively, with only a minor modification due to the fact that the two ferromagnetic materials in our junctions are not the same.  Accordingly, we use $(d_\mathrm{Ni}/\xi_\mathrm{Ni} \pm d_\mathrm{NiFe}/\xi_\mathrm{NiFe})$ in place of $(d_{F1} \pm d_{F2})/\xi_F$ in the Crouzy equations.  For each data set in Fig. 3, we fit both the P and AP states simultaneously, with the only free parameters being an overall magnitude, $I_{c0}R_N$, and the characteristic length scales for the two F materials: $\xi_\mathrm{Ni}$ and $\xi_\mathrm{NiFe}$.  The results are shown as the solid black and red lines in those figures, for the P and AP states, respectively.  The fit lines do a decent job of describing the trends in the data.  (The fits in Fig. 3(b) appear to be low because the error bars on the higher data points are generally larger than those on the lower data points.)  The fit parameters are listed in the first two lines of Table I.

\begin{table}[ptbh]
\begin{tabular}
[c]{|c|c|c|c|}\hline
Data Set & $I_{c0}R_N$ ($\mu V$) & $\xi_\mathrm{Ni}$ (nm)& $\xi_\mathrm{NiFe} (nm)$\\
\hline
Ni(2.0) zero-field data & 563 & 0.612 & 0.657\\
Ni(2.0) Fraunhofer fits & 493 & 0.614 & 0.656\\
Ni(1.6) zero-field data & 388 & 0.527 & 0.640\\
\hline
\end{tabular}
\newline \caption{First two lines: fit parameters for the lines shown in Figs. 3a) and 3b), based on Eqns. (17) and (20) in Ref. \onlinecite{Crouzy:07}. Last line: fit parameters for the lines shown in Fig. 5.  (The three digits of precision are not meant to indicate statistical significance,\cite{Uncertainties} but are included in case a reader wants to reproduce the fit curves in the figures.)} \label{FitParams}
\end{table}

Before discussing the significance of the fit parameters shown in Table I, we make two additional observations. First, the Crouzy theory implicitly assumes that the length scales governing the 0-$\pi$ oscillations and the exponential decay of the supercurrent with F-layer thicknesses are the same, whereas our data on S/F/S junctions containing only NiFe show a difference between those lengths.\cite{Glick:17}  Given the rather small range of NiFe thicknesses in our junctions, this constraint in the Crouzy theory should not present a major problem.  Second, it is instructive to compare our results with those of Baek \textit{et al}.\cite{Baek:14,Baek:17}  Those authors studied spin valve junctions containing a Ni layer of variable thickness and a Nb-doped NiFe layer of fixed thickness.  They were able to fit their data\cite{Baek:14} using the purely ballistic theory of Buzdin \textit{et al.}\cite{Buzdin:82} for ballistic S/F/S junctions, with different thickness offsets for the P and AP states to account for the electron pair phase accumulated in the NiFeNb layer.  They noted later,\cite{Baek:17} however, that the location of the first 0-$\pi$ transition as a function of Ni thickness shifted substantially from 0.9 nm to 1.5 nm after the NiFeNb layer was added to the junction.  Such a shift cannot be explained by the purely ballistic theory, and those authors speculated that the NiFeNb causes a partial crossover to diffusive transport in the junction.  We believe that our samples are in a similar crossover regime between ballistic and diffusive transport; unfortunately, there is no theory for the supercurrent in spin-valve Josephson junctions in such a crossover regime.  Nevertheless, one might ask why we are unable to fit the ballistic theory to our data.  There are some important differences between the Baek work and ours. Those authors vary the Ni thickness in their series of junctions; the changing Ni thickness causes only a slow algebraic decay in the overall magnitude of $I_{c}R_N$ across the series.\cite{Baek:17}  In contrast, we vary the NiFe thickness in our series of junctions; the changing NiFe thickness causes a much faster exponential decay in the overall magnitude of $I_{c}R_N$ across the series.\cite{Glick:17}  In addition, most of the electron-pair phase accumulation in the Baek samples occurs in the Ni layer, while the weakly-magnetic NiFeNb layer only shifts the positions of the 0-$\pi$ transitions slightly to the left or right along the Ni thickness axis.  In our work, the strong magnet NiFe contributes substantially to the accumulated phase shift.  As we will discuss below, the fits to our data suggest that the phase shift acquired by an electron pair traversing 1.8 nm of NiFe is nearly as large as that acquired traversing 2.0 nm of Ni, so we must treat the Ni and NiFe layers on an equal footing.

Returning to the fit parameters shown in Table I, the value we find of $\xi_\mathrm{Ni} = 0.61$nm is significantly smaller than Baek's value of 0.95 nm.  That is the first indication that we should beware of over-interpreting the fits shown in Fig. 3.  Proceeding nonetheless, we would say that the electron pair phase shift through the 2.0-nm-thick Ni layer is $d_\mathrm{Ni}/\xi_\mathrm{Ni}$ = 3.3 = 1.04~$\pi$.  The value of $\xi_\mathrm{NiFe} = 0.66$nm in the Table is not far from the value of 0.58 nm obtained in our recent study of S/F/S junctions containing only NiFe,\cite{Glick:17} although we note that only a single 0-$\pi$ transition was unambiguously observed in that study, so the value of $\xi_F$ was somewhat uncertain.  Using the value of $\xi_\mathrm{NiFe}$ from the fit, we calculate that the electron pair phase accumulation through the NiFe layers in our samples, $d_\mathrm{NiFe}/\xi_\mathrm{NiFe}$, ranges from 1.7 to 2.7 as $d_\mathrm{NiFe}$ ranges from 1.1 to 1.8 nm.  According to the fit, then, the net phase shift in the AP state in the sample with the thickest NiFe layer is only about $3.3-2.7=0.6$ radians.  This observation highlights why a purely ballistic theory could not fit our data, as it would predict an extremely large value of $I_cR_N$ in the AP state of those junctions.  Regarding the ``phase diagram" of the junctions, The Crouzy model provides not only the magnitude of the critical current, but also its sign, with a negative sign corresponding to the $\pi$ state of the junction.  (The data and fits shown in the figures represent the absolute values of $I_cR_N$, since our measurements do not provide information on the junction phase state.)  The fits of the Crouzy model to the data shown in Fig. 3 imply that the junctions in the P state are $\pi$-junctions for $d_\mathrm{NiFe} < 1.47$nm, and switch to the 0-state for $d_\mathrm{NiFe} > 1.47$nm.  In the AP state, the junctions are 0-junctions for nearly the whole series except for those with $d_\mathrm{NiFe} = 1.1$nm.  Hence the NiFe thickness range over which a single junction could be switched between the 0 and $\pi$ states is limited to 1.14 nm$ < d_\mathrm{NiFe} < 1.47$nm.

\begin{figure}[tbh!]
\begin{center}
\includegraphics[width=3.0in]{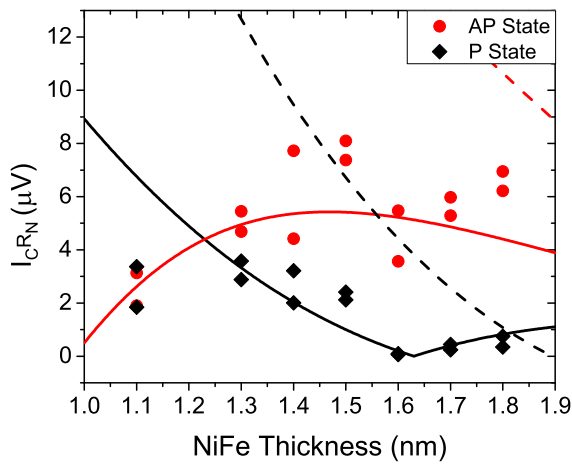}
\end{center}
\caption{(color online) Critical current times normal state resistance of Josephson junctions with thinner Ni(1.6) layer vs NiFe layer thickness.  Red circles represent the AP state while black diamonds represent the P state.  The solid lines are fits to Eqns. (17) and (20) in Ref. \onlinecite{Crouzy:07} with $\xi_\mathrm{Ni}=0.531$nm and $\xi_\mathrm{NiFe}$ kept at 0.656nm as described in the text.  The dashed lines come from the same equations but with the fit parameters determined from the original data set of samples with Ni(2.0).  The AP-state critical current is grossly overestimated, with the curve barely visible in the upper-right portion of the graph.}\label{IcRn16}
\end{figure}

A true test of a theory is its predictive ability.  Based on the results described above, we fabricated and measured a second series of junctions with a Ni thickness of 1.6 nm, and with the same range of NiFe thicknesses, 1.1 - 1.8 nm.  Due to the smaller Ni thickness, we expected to find a broader range of NiFe thicknesses over which the P state and AP state correspond to different phase states of the junction.  We also expected to find much larger values of $I_cR_N$ in the middle of the NiFe thickness range.  Unfortunately the Fraunhofer patterns of the samples in this second series were not quite as nice as those in the first series, so we did not attempt to fit the P-state and AP-state Fraunhofer patterns to Eqn.~(\ref{eqn:FraunhoferAiryFit}).  Nevertheless, we extracted values of $I_cR_N$ from the zero-field data, which are shown in Fig. 5.  Compared to the AP-state data shown in Fig. 3(b), the AP-state data in Fig. 5 do indeed exhibit somewhat larger values of $I_cR_N$, with the location of the nearest zero in $I_cR_N$ pushed off to the left -- beyond the minimum value of $d_\mathrm{NiFe}$=1.1 nm in the series.  The P-state data in Fig. 5 differ less from those in Fig. 3, but it appears that the zero in $I_cR_N$ has moved a bit to the right, perhaps even beyond 1.6 nm.  So in a qualitative sense, the data correspond to our expectations.  Trying to fit the Crouzy model to these data, however, leads to a gross inconsistency.  The theoretical curves corresponding to the fit parameters in the first two rows of Table I, but with the Ni thickness set at 1.6 nm rather than 2.0 nm (dashed lines in Fig. 5), drastically overestimate the values of $I_cR_N$ in both states, but especially in the AP state.  If instead we perform a least-squares fit to Ni(1.6) data set, we obtain the fit parameters shown in the last line of Table I.  The value of $\xi_\mathrm{Ni}$ that gives the best fit to the data has decreased from 0.61 to 0.53 nm, whereas the value of $\xi_\mathrm{NiFe}$ has stayed about the same.  The overall amplitude, $I_{c0}R_N$, has also decreased significantly.  Let us imagine that the properties of the Ni layer have somehow changed dramatically when its thickness was reduced from 2.0 to 1.6 nm.  Then we can ask what happens if we keep the original value of $\xi_\mathrm{NiFe}$=0.656 nm while letting $\xi_\mathrm{Ni}$ vary.  The result is a fit that is statistically indistinguishable from the least-squares fit, with $\xi_\mathrm{Ni}$=0.531 nm only slightly different from the value $\xi_\mathrm{Ni}$=0.527 nm shown in the last line of the table.  This modified fit is shown as the solid lines in Fig. 5, and it fits the data rather well.  But the inescapable conclusion of this fitting exercise if that no single set of fit parameters gives satisfactory fits to both the Ni(2.0) and Ni(1.6) data sets.

\section{Conclusions}

In conclusion, we have performed a comprehensive study of spin-valve Josephson junctions with a hard magnetic layer of Ni with fixed thickness of 2.0 nm and a soft magnetic layer of NiFe with variable thickness between 1.1 and 1.8 nm.  We have also performed a limited study of similar junctions with a slightly thinner 1.6-nm Ni layer.  The junctions all exhibit clear switching between the parallel and antiparallel magnetic states.  The field shifts of the Fraunhofer patterns in both the P and AP states generally follow the expected trend based on the net magnetic moments in the junctions, although there is significant scatter in the data.

The most important result of this work is the preliminary mapping of the phase diagram for spin-valve junctions whose ground-state phase difference can be controllable toggled between 0 or $\pi$, based on the locations where the critical current passes through a minimum.  With the Ni thickness set at 2.0 nm, $I_c$ in the AP state appears to pass through a minimum when the NiFe thickness is in the vicinity of 1.1 nm.  In the P state, the location of the minimum appears to be at a NiFe thickness of about 1.5 nm.  Theoretical modeling suggests that the P states correspond to $\pi$-junctions and the AP states correspond to 0-junctions for NiFe thicknesses between those two values.  In the second set of samples with $d_\mathrm{Ni}$ = 1.6 nm, the minima in $I_cR_N$ appear to move outward, so that the range of NiFe thicknesses where a junction could be switched between its 0 and $\pi$ states increases.  Confirmation of these statements would require performing phase-sensitive measurements, as we did a year ago with junctions containing Ni(1.2) and NiFe(1.0) hard and soft layers.\cite{Gingrich:16} If we apply the theoretical modeling presented here to those samples in our previous work,\cite{Gingrich:16} we find that it correctly postdicts both the fact that the AP and P magnetic states corresponded to the 0 and $\pi$ phase states of the junctions, respectively, and that the magnitude of the critical current was larger in the AP state than in the P state.  But the model grossly overestimates the magnitudes of the critical currents in both states.

While the theoretical community has made great strides in understanding many aspects of ferromagnetic Josephson junctions, the lack of a suitable theoretical formula to describe these complex spin-valve samples quantitatively remains a hindrance to the development of practical devices.  Further theoretical and experiment work is badly needed to move this field onto a firmer footing where device design can be optimized through a synergy of theory and experiment.

Acknowledgements: We are grateful to V. Aguilar for his assistance with the measurement software.  We acknowledge fruitful conversations with F.S. Bergeret, A.I. Buzdin, E.C. Gingrich, A.Y. Herr, D.L. Miller, O. Naaman, N. Newman, and N.D. Rizzo. We also thank B. Bi for technical assistance, and the use of the W.M. Keck Microfabrication Facility. This research is supported by the Office of the Director of National Intelligence (ODNI), Intelligence Advanced Research Projects Activity (IARPA), via U.S. Army Research Office contract W911NF-14-C-0115. The views and conclusions contained herein are those of the authors and should not be interpreted as necessarily representing the official policies or endorsements, either expressed or implied, of the ODNI, IARPA, or the U.S. Government.


\begin{thebibliography} {99}
\bibitem{IVcurvenote} For junctions with $I_c$ less than about 10 $\mu A$, there is noticeable rounding of the I-V curves due to instrumental noise.  Such data can be fit more accurately using the Ivanchenko-Zil'berman theory, as described in ref. [17].  There was no need to carry out that more elaborate fitting procedure for the data in this work.
\bibitem{badmilling} It is tempting to speculate that there was an error in the ion milling step for that chip, resulting in magnetic layers that were not completely milled.  But the Fraunhofer patterns do exhibit sharp jumps near zero field associated with switching of the NiFe layer, which appears to rule out this ``in sufficient milling" hypothesis.
\end{thebibliography}
\end{document}